# Entangled Photon Holes

J.D. Franson

*Johns Hopkins University Applied Physics Laboratory, Laurel, MD 20723*

Most experimental demonstrations of entanglement require nonclassical states and correlated measurements of single-photon detection events. It is shown here that entanglement can produce a large decrease in the rate of two-photon absorption for a classical input state that can be observed using classical detectors. These effects can be interpreted as being due to the creation of entangled photon holes that are somewhat analogous to the holes of semiconductor theory.

Entanglement is one of the most fundamental properties of quantum systems and it plays a major role in quantum information processing, for example. Here we show that a classical input state incident on a three-level atomic medium will undergo two-photon absorption [1-13] at a rate that is greatly reduced by the generation of entangled photon holes that are somewhat analogous to the holes of semiconductor theory. The effects of entanglement can then be observed using a classical detector, such as an intensity meter. The entangled photon holes can also violate Bell's inequality if single-photon detectors are used.

Many nonclassical features of two-photon absorption have already been described [3-12], including an enhanced rate of two-photon absorption when the incident photons are entangled [3, 8-9, 12]. The pairs of photons from parametric down-conversion are known to have been emitted at nearly the same time, but that time is completely uncertain in the quantum-mechanical sense, as illustrated in Fig. 1a [14]. The fact that the photons are incident on any given atom at the same time while their total energy is still well defined gives rise to an increase in the rate of two-photon absorption, which can be linearly dependent on the intensity of the incident beam [8-9, 12].

The situation of interest here is essentially the inverse of parametric down-conversion, as illustrated in Fig 1b. In the limit of large detunings, three-level atoms will absorb pairs of photons at very nearly the same time, producing a decrease in the probability amplitude for both of the photons to be at the same location. In analogy with the holes of semiconductor theory, the reduced probability amplitudes of Fig. 1b can be viewed as entangled photon holes in an otherwise constant background. Entanglement of this kind can reduce the rate of two-photon absorption to a level that is substantially less than that of classical or semiclassical theory. Roughly speaking, the magnitude of the dips in the probability amplitude will continue to increase until there is no significant probability amplitude for two photons to be found at the same location.

The state vectors corresponding to the probability amplitudes of Figs 1a and 1b cannot be written as the product of two single-particle states and both systems are thus in an entangled state. One way to demonstrate the entanglement is by showing that Bell's inequality can be violated, as will be done later in the paper after we first consider the macroscopic effects of the entangled photon holes.

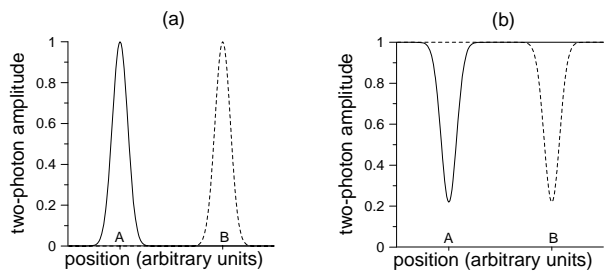

FIG. 1. (a) Two-photon probability amplitude from parametric down-conversion, where the photons are known to be at the same location (to within the bandwidth of the source) with an equal probability amplitude for all locations, such as points A and B; (b) Two-photon probability amplitude for the entangled photon holes produced by two-photon absorption, where there is an equal probability amplitude for two photons to have been removed from any location.

Two-photon absorption is often treated semiclassically [1, 13], by using one or two quantized field modes [5-7], or to lowest order in perturbation theory for a continuum of field modes [4, 8-9, 11-12, 15]. Although analytic results can be obtained using those approximations, the effects of interest here require a multi-mode calculation performed to all orders. In the absence of an analytic solution, numerical methods were used to integrate Schrodinger's equation as the incident photons propagated past a series of three-level atoms. The field was assumed to propagate in only one longitudinal dimension $x$, which would be a reasonable approximation for light propagating in a single-mode optical fiber, for example.

Periodic boundary conditions were assumed with periodicity $L$, which determines the spacing between the allowed wave vectors $k$. The incident photons were assumed to be non-degenerate with central wave vectors $k_{01}$ and $k_{02}$. Only a limited number $n$ of modes for each photon could be included in the numerical calculations, and wave vectors outside of a band $\Delta k$ centered about $k_{01}$ and $k_{02}$ were neglected. This limits the spatial resolution of the numerical results but it is sufficient to demonstrate the effects of interest, as will be discussed below. It was further assumed that the energy of photon 1 differed from that of the first excited atomic state by $\hbar\delta\omega$, where $\delta\omega$ is the detuning, so that photon 1 would be absorbed first in a virtual transition to the upper atomic level.

The incident fields were assumed to be weak coherent states with negligible probability that more than two photons will be incident on the atomic medium. The only contribution to the two-photon absorption rate occurs when a single photon is present in each beam, so that we only need to consider the terms in the initial state vector that correspond to single-photon number states for each photon. The incident photons were assumed to be in gaussian wave packets with width $\delta k \ll \Delta k$ in momentum space, and the spacing between the atoms was taken to be sufficiently large that the photons only interacted with one atom at a time, which further simplified the calculations.

With these approximations, there were $n^2$ terms in the state vector corresponding to two photons with wave vectors $k_1$ and $k_2$, $n$ terms corresponding to a single photon and the atom in the first excited state, and one additional state with the atom in the upper level. This gives a total of $n^2 + n + 1$ basis states in Hilbert space, and the Hamiltonian describing this system contained more than $n^4$ terms (most of which were zero). Schrodinger's equation for this system was integrated numerically using Mathematica, and it was found that the required execution time did scale as $n^4$ even though the Hamiltonian was sparse. The necessary calculations could be performed for a given set of parameters in four hours for $n = 50$, which was taken to be the baseline for all of the calculations reported here. A subset of the calculations were repeated using $n = 100$, which gave nearly identical results and justified the use of $n = 50$ for the remaining calculations.

In this basis, the Hamiltonian for the system can be written as

$$\hat{H} = \sum_{k_1}(\hat{a}_{k1}^\dagger \hat{a}_{k1} + 1/2)\hbar\omega_1 + \sum_{k_2}(\hat{a}_{k2}^\dagger \hat{a}_{k2} + 1/2)\hbar\omega_2$$
$$+ E_1 \hat{\sigma}_{G1}^\dagger \hat{\sigma}_{G1} + E_2 \hat{\sigma}_{12}^\dagger \hat{\sigma}_{12} + \sum_{k_1} M_1 \hat{\sigma}_{G1}^\dagger \hat{a}_{k1} e^{ik_1 x} \quad (1)$$
$$+ \sum_{k_2} M_2 \hat{\sigma}_{12}^\dagger \hat{a}_{k2} e^{ik_2 x} + h.c.$$

Here $E_1$ and $E_2$ are the energies of the relevant atomic states, the operators $\hat{a}_{k1}^\dagger$ and $\hat{a}_{k2}^\dagger$ create a photon with wave vector $k_1$ or $k_2$, $\hat{\sigma}_{G1}^\dagger$ produces a transition from the atomic ground state to the first excited state, $\hat{\sigma}_{12}^\dagger$ produces a transition to the second excited state, $M_1$ and $M_2$ are the corresponding matrix elements, and $x$ is the position of the atom. The value of $E_2$ was adjusted to be on resonance while $\omega_1(k_1)$ and $\omega_2(k_2)$ were chosen to ensure that the group velocities of the two photons were the same. This Hamiltonian does not include any loss or decoherence of the excited atomic states, which is a valid approximation when the transit time of the photon wave packets through the location of an atom is small compared to the decoherence time. Schrodinger's equation for this Hamiltonian corresponded to a set of $n^2 + n + 1$ coupled differential equations that were solved numerically.

It is convenient to define the factors $F_1$, $f_1$, and $g_1$ by the relations

$$\Delta k_1 = F_1 k_{01}$$
$$\delta k_1 = f_1 k_{01} \quad (2)$$
$$M_1 = g_1 \frac{\hbar\omega_{01}}{2}.$$

The final equation defines the coupling constant $g_1$ in such a way that $g_1 = 1$ would correspond to a Rabi frequency of $\omega_{01}$. Similar parameters were defined for photon 2, with $F_2 = F_1$ and $f_2 = f_1$. The value of $k_{01} = 2\pi/\lambda_0$ was chosen to correspond to a central wavelength $\lambda_0$ of 1 $\mu m$. Although calculations were performed for a range of parameters, the results shown here all correspond to $F_1 = 0.01$, $f_1 = 0.001$, $g_1 = 0.0035$, $g_2 = 0.00071$, and $\delta\omega = 0.1\omega_{01}$, unless otherwise noted. The values of $g_1$ and $g_2$ were chosen to give roughly 2% two-photon absorption per atom in order to reduce the number of atoms required to demonstrate the effects of interest. These matrix elements are larger than would be expected for a typical atomic transition, but this approach was necessary in order to limit the computer execution time and it does not affect the nature of the results.

Fig. 2a shows the initial intensity distribution $I_1(x)$ of the field associated with photon 1 while Fig. 2b shows the probability $P_2(s)$ of detecting two photons separated by a distance $s$. The single-photon intensity $I_1(x)$ was calculated by tracing over the photon-2 components while $P_2(s)$ was given as usual by

$$P_2(s) = \eta \int dx_1 \left\langle \hat{E}_1^-(x_1)\hat{E}_2^-(x_1+s)\hat{E}_2^+(x_1+s)\hat{E}_1^+(x_1) \right\rangle (3)$$

where $\eta$ is a constant associated with the detection efficiency and time window.

It was assumed that the atoms were located along the path of the photons with a separation of 1 mm,

which is much longer than the width of the wave packets of Fig. 2a. The initial state was then propagated for a time interval $\Delta t$ during which the photons interacted with one or more atoms. The single-photon intensity and coincidence detection probability are shown in Figs. 2c and 2d for the case in which the photon wave packets had propagated a distance of 5 mm and interacted with 5 atoms. It can be seen that the single-photon intensity $I_1(x)$ has nearly the same shape that it did initially while the coincidence counting probability $P_2(s)$ shows a small dip centered on $s = 0$ as would be expected.

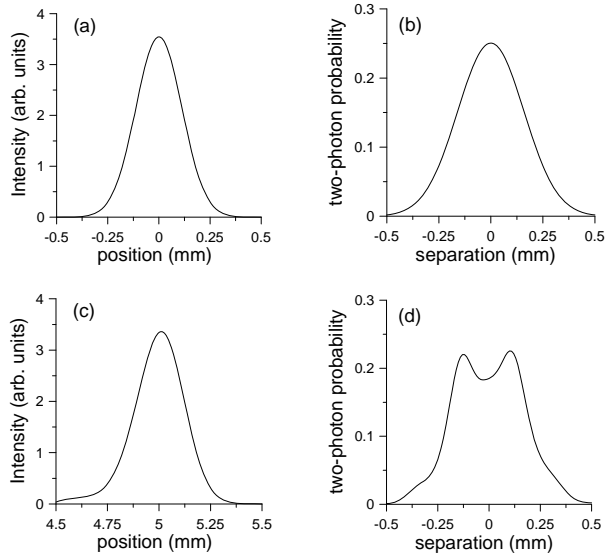

FIG. 2. Single-photon intensity (a) as a function of position and two-photon detection probability (b) as a function of separation in the initial state. The corresponding results after passing through five atoms are shown in (c) and (d).

The probability $P_I$ of remaining in the initial two-photon state is plotted in Fig. 3a as a function of the number of atoms with which the photons have interacted. It can be seen that the two-photon absorption rate decreases for larger numbers of atoms and that $P_I$ approaches a plateau after ~ 20 atoms, which is consistent with the expected effects of the entangled photon holes. Additional calculations were performed using twice the bandwidth $\Delta k$, which resulted in a narrower coincidence dip and a plateau at a larger value of $P_I$ as would be expected.

A semiclassical treatment of two-photon absorption for $\omega_{01} \sim \omega_{02}$ gives [1]

$$\frac{dI_1}{dt} = \frac{dI_2}{dt} = -\alpha I_1 I_2. \qquad (4)$$

Here $\alpha$ is a constant that can be derived from the atomic matrix elements and $I_1$ and $I_2$ are the intensities of the two beams. This set of equations was also integrated numerically using a value of $\alpha$ that was chosen to give a two-photon absorption rate similar to that of the earlier calculations. The integrated intensity (total energy) is plotted as a function of time in Fig. 3b. It can be seen that the semiclassical rate of two-photon absorption also deviates from a simple exponential due to the nonlinear nature of Eq. (4), but it does not approach a plateau like the data of Fig. 3a. Thus the decrease in the rate of two-photon absorption due to the entangled photon holes is inconsistent with semiclassical theory.

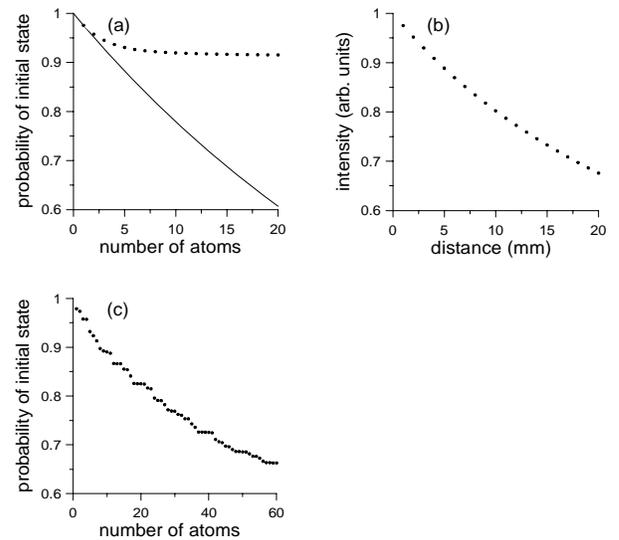

FIG. 3. (a) The dots indicate the probability of remaining in the initial two-photon state after passing through $n_A$ atoms, while the line represents an exponential decay curve fitted to the first two data points. (b) The integrated intensity as a function of propagation distance as calculated using semiclassical theory. (c) The probability of remaining in the initial two-photon state for the case in which the photons are traveling in opposite directions. The fluctuations in these data are due to the random choice of the positions of the atoms.

We have previously suggested that quantum logic operations could be performed using the quantum Zeno effect imposed by strong two-photon absorption [15], and it is apparent that entangled photon holes may have an adverse effect on the operation of devices of that kind. These difficulties can be avoided if the photons propagate in opposite directions, in which case the photon holes will travel away from each other and the probability of finding two photons in the same location will not be prematurely depleted. This can be seen in Fig. 3c, where it was assumed that the photons propagated in opposite directions around a ring with a 1 mm circumference, with the atoms located in randomly-chosen positions. Although the two-photon absorption no longer saturates as before, the data does depart from a true exponential decay due to the effects of dispersion on the wave packet of photon 1. Dispersion also plays a role for the data of Fig. 3a for larger numbers of atoms.

The decreased two-photon absorption rate does not violate Bell's inequality in any obvious way, and one might ask whether or not it really depends on entanglement or if it could be described by a mixed state instead. The final probability amplitude for an atom to be in the upper state corresponds to a coherent sum of the contributions from all of the possible locations of the entangled holes, and different results would be obtained depending on the nonlocal phase properties of the entangled state. Thus the fourth-order coherence properties of the entangled photon holes is just as important here as it is for the nonlocal interference effects described below.

We now return to the entangled nature of the photon holes and show that they can violate Bell's inequality if single-photon detectors are used. First consider the case of entangled pairs of photons from parametric down-conversion, which I discussed in an earlier paper [14]. We assume that the photons propagate through two distant Mach-Zehnder interferometers containing a long path $L$ and a short path $S$, with phase shifts $\phi_1$ and $\phi_2$ in the two longer paths. If we only consider coincident events in which the photons arrive at two single-photon detectors at the same time, there will be no contribution from events $L_1S_2$ or $S_1L_2$ in which the photons travel paths of different lengths. Quantum interference between the probability amplitudes for $S_1S_2$ and $L_1L_2$, where both photons travel the longer or the shorter paths, produces a coincidence counting rate proportional to $\cos^2[(\phi_1+\phi_2)/2]$ (aside from a constant phase factor), which violates Bell's inequality [14].

Bell's inequality can be violated in a similar way by the entangled photon holes of Fig. 1b if the initial state corresponds to two weak coherent states at two different frequencies $\omega_1$ and $\omega_2$. The probability that more than one photon will be present in either beam is assumed to be negligible and the dips in the two-photon probability amplitude are assumed to go to zero. Using the same dual-interferometer arrangement, there will now be no contribution from the $S_1S_2$ and $L_1L_2$ events, since no photons are emitted from the source (two-photon absorbing medium) at the same time. Quantum interference between the probability amplitudes corresponding to $L_1S_2$ and $S_1L_2$ will produce a coincidence rate proportional to $\cos^2[(\phi_1-\phi_2)/2]$, which also violates Bell's inequality. The sum of the frequencies of the detected photons is well defined, just as in the original case of down-converted photons, but here the coherence comes directly from the input state. A more detailed derivation of these results using the second-quantized approach of Ref. 14 will be submitted for publication elsewhere.

The analogy between these effects and the holes of semiconductor theory is obviously limited by the fact that photons are bosons and not fermions. As a result, the background states are occupied with a probability amplitude much less than one in the examples of interest here, in contrast to the unit probability amplitudes for fermions below the Fermi level. Nevertheless, the situations are analogous in the sense that there is "an unoccupied space" in an otherwise constant background, which is the conventional definition of a hole. The concept of holes is of obvious use in understanding the nature of excitons in semiconductor theory, for example, and the notion of entangled photon holes has already been useful in predicting and understanding the phenomena described here. It is hoped that the notion of photon holes will be of wider use in other areas of quantum optics as well.

I would like to thank Todd Pittman for his comments on the paper. This work was supported by DTO, ARO, and IR&D funds.